\documentclass{article}
\usepackage{spconf,amsmath,epsfig}
\usepackage{multirow}
\usepackage{xcolor}
\usepackage{graphicx,epstopdf}
\usepackage{amsmath,amsthm,amssymb,amsfonts}
\usepackage{tabularx,pdfcomment}
\usepackage{mathrsfs}
\usepackage{subfig}
\usepackage{hyperref}
\usepackage{url}
\usepackage{cases}
\usepackage{bbm}
\usepackage{multicol}
\usepackage{bm}
\usepackage{algorithmic}
\usepackage{algorithm}
\theoremstyle{plain} 
\theoremstyle{plain} \newtheorem{definition}{Definition}
\theoremstyle{plain} 
\theoremstyle{plain} 
\theoremstyle{plain} 
\theoremstyle{plain}

\DeclareMathOperator*{\argmin}{arg\,min}

\newcommand{\norm}[1]{\left\lVert#1\right\rVert}

\newcounter{RomanNumber}

\begin{document}\sloppy

% Title.
% ------
\title{Multidimensional Data Tensor Sensing for RF Tomographic Imaging}
%
% Single address.
% ---------------
\name{Tao Deng$^{1}$, Xiao-Yang Liu$^{2}$, Feng Qian$^{1}$ and Anwar Walid$^{3}$}
\address{$^1$School of Communication and Information Engineering,\\
	 University of Electronic Science and Technology of China\\
	$^2$Dept. of Electrical Engineering, Columbia University\\
	$^3$Bell Laboratories\\}

\maketitle

\begin{abstract}
Radio-frequency (RF) tomographic imaging is a promising technique for inferring multi-dimensional physical space by processing RF signals traversed across a region of interest. However, conventional RF tomography schemes are generally based on vector compressed sensing, which ignores the geometric structures of the target spaces and leads to low recovery precision. The recently proposed transform-based tensor model is more appropriate for sensory data processing, as it helps exploit the geometric structures of the three-dimensional target and improve the recovery precision. In this paper, we propose a novel tensor sensing approach that achieves highly accurate estimation for real-world three-dimensional spaces. First, we use the transform-based tensor model to formulate a tensor sensing problem, and propose a fast alternating minimization algorithm called Alt-Min. Secondly, we drive an algorithm which is optimized to reduce memory and computation requirements. Finally, we present evaluation of our Alt-Min approach using IKEA 3D data and demonstrate significant improvement in recovery error and convergence speed compared to prior tensor-based compressed sensing.
\end{abstract}
\begin{keywords}
RF Tomographic Imaging, Transform-based Tensor, Tensor Sensing, Alternating Minimization
\end{keywords}
\section{Introduction}
\label{sec:intro}
RF Tomographic imaging \cite{kanso2009compressed, liutkus2014imaging} is an inference technique which can be used for remotely learning the locations and the shapes of objects, as depicted in Fig. 1. RF tomographic imaging can be applied in smart buildings and spaces applications \cite{wilson2010radio}, and in emergencies, rescue operations and security breaches \cite{matsuda2017multi}, since the objects being imaged need not carry an electronic device or a cellphone.

There have been considerable works on reconstruction methods for RF tomographic imaging which can be categorized into vector-based and tensor-based methods. In \cite{kanso2009compressed, Mostofi2011Compressive}, vector-based compressed sensing approaches for RF tomographic imaging are proposed. These vector-based methods are designed to infer two-dimensional spaces, and they do not have the ability to estimate three-dimensional spaces since spatial structures of the data are inherently ignored. Reference \cite{matsuda2017multi} proposes the tensor-based compressed sensing algorithm using tensor nuclear norm (TNN) \cite{li2013generalized}, which extends the RF tomographic imaging problem to three-dimensional case. However, it requires computing tensor singular value decomposition (t-SVD), which leads to high computational complexity, especially for large scale tensors. Moreover, the recovery error of this approach is relatively high, and the data size that can be handled is too small for realistic scenarios. In this paper, we aim to exploit the spatial structures of three-dimensional spaces for more efficient and accurate inference.

\begin{figure}[t]
	\centering
	\includegraphics[totalheight=5.5cm]{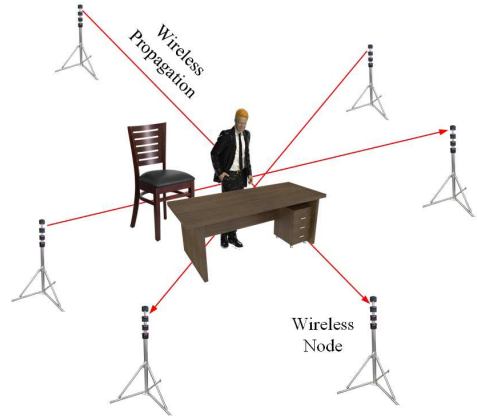}
	\caption{An illustration of RF tomographic imaging network. Each node broadcasts to the others, creating many projections that can be used to reconstruct objects inside the network area.}
	\label{fig1}
	\vspace{-0.15in}
\end{figure}

In order to represent three-dimensional spatial structures effectively, we adopt the recently proposed transform-based tensor model \cite{liu2017fourth}, which has the following advantages: (i) compared with other tensor models, only real-valued fast transforms are involved, so it is appropriate for the received signal strength (RSS) data; (ii) unlike other existing tensor models, the transform-based tensor model allows for diverse sampling strategies. On the other hand, we decompose the loss field tensor into the product of two tensors of small sizes, so that we only need to iteratively update two small tensors, which is much more effective than prior tensor-based compressed sensing.
 
In this paper, we propose a novel RF tomographic imaging scheme using tensor sensing to estimate the three-dimensional spaces. We formulate the RF tomographic imaging as a tensor sensing problem in a transform domain, and propose a novel Alternating-Minimization (Alt-Min) algorithm, whose implementation is optimized for memory consumption and computation speed. We apply our algorithm to the IKEA 3D datasets \cite{lpt2013ikea}, and demonstrate significantly lower recovery error and required number of iterations, compared with the approach in \cite{matsuda2017multi}. 
 
The remainder of the paper is organized as follows: system model and problem formulation are given in Section II. Section III presents the solution algorithm, implementation and algorithm optimization. The evaluation results are presented in Section IV. Finally, we conclude this paper in Section V.

%%%%%%%%%%%%%%%%%%%%%%%%%%%%%%%%%%%%%%%%%%%%%%%%%%%%%%

\section{Problem Statement and Formulation}

We first review the RF tomographic imaging problem \cite{kanso2009compressed,wilson2010radio,adib20143d}, then model the shadowing loss data as a transform-based tensor, and formulate the RF tomographic imaging task as a tensor sensing problem.

\noindent\textbf{Notations-} We use lowercase boldface letter $\bm{x}\in \mathbb{R}^{N_1}$ to denote a vector, uppercase boldface letter $\bm{X}\in \mathbb{R}^{N_1 \times N_2}$ to denote a matrix, and calligraphic letter $\mathcal{X} \in \mathbb{R}^{N_1 \times N_2 \times N_3}$ to denote a tensor. Let $[k]$ denote the set $\{1,2,\cdots,k\}$.
%%%%%%%%%%%%%%%%%%%%%%%%%%

\subsection{RF Tomographic Channel Model}
We consider the RF tomographic imaging with the space of interest being represented as a 3D tensor in Cartesian coordinates. A set of RF signal nodes are uniformly deployed around the sides of the ``tensor", forming a complete tomography network. Any pair of nodes can establish a unique link. The RF signal on a given link suffers from path loss, which consists of three parts: (i) shadowing loss due to obstructions; (ii) distant-dependent large-scale path loss; (iii) non-shadowing loss due to multipath \cite{kanso2009compressed}. We use $V$ to represent the node set, and assume that $v_i\in V$ is a transmitter and $v_j\in V(i\ne j)$ is a receiver. Let $\bm{P}_{ij}$ be the received power at $v_j$, then we can obtain the following power equation:
\begin{equation} 
\begin{split}
\bm{P}_{ij}&=\bm{P}_{t}-\overline{\bm{P}}(d_{ij})-\bm{Z}_{ij},\\
\bm{Z}_{ij}&=\bm{Z}_{ij}^{\left( 1\right)} + \bm{Z}_{ij}^{\left( 2\right)},
\end{split}
\end{equation}
where $\bm{P}_{t}$ is the transmitted power, $d_{ij}$ is the distance between node $v_i$ and $v_j$, and $\overline{\bm{P}}(d_{ij})$ represents the corresponding large-scale path loss of the link. Let $\bm{Z}_{ij}$ be the total fading loss that involves the non-shadowing fading loss $\bm{Z}_{ij}^{\left( 2\right)}$ and the shadowing loss $\bm{Z}_{ij}^{\left( 1\right)}$. 

% Specifically, $\overline{\bm{{P}}}(d_{ij})$ is given as follows:
% \begin{equation} 
%  \overline{\bm{P}}(d_{ij})=10\rho\log{\left(d_{ij}\right)}+\lambda
% \end{equation}
% where $\rho$ is the path loss exponent \cite{matsuda2017multi}, and $\lambda$ is a correction parameter. 

Assuming that the space of interest $\mathcal{X}\in\mathbb{R}^{N_1 \times N_2 \times N_3}$ is divided into a set of three-dimensional voxels of the same size. We use $\Delta(n_1,n_2,n_3)\in\mathbb{R}^{3}$ $(n_1\in[N_1]$, $n_2\in[N_2]$, $n_3\in[N_3])$ to denote a voxel at the corresponding coordinate. When the RF signal propagates through the voxels, the total value of shadowing loss $\bm{Z}_{ij}^{\left( 1\right)}$ is equivalent to the total attenuation that arises in each single voxel. Moreover, we assume that the internal medium of each voxel is homogeneous, and the power attenuation coefficient is a constant value $\mathcal{X}(n_1,n_2,n_3)$ within $\Delta(n_1,n_2,n_3)$ \cite{matsuda2017multi}. Given these definitions, the shadowing loss of a unique $\text{link}(i,j)$ between $v_i$ and $v_j$ can be formulated as:
 \begin{equation} 
\bm{Z}_{ij}^{\left( 1\right)} = \sum_{n_1,n_2,n_3} {\mathcal{D}_{ij}(n_1,n_2,n_3)\mathcal{X}(n_1,n_2,n_3)},
\end{equation} 
where $\mathcal{D}_{ij}(n_1,n_2,n_3)$ represents the overlapped distance between the $\text{link}(i,j)$ and voxel $\Delta(n_1,n_2,n_3)$. The non-shadowing fading loss $\bm{Z}_{ij}^{\left( 2\right)}$ is assumed to be a stationary Gaussian process with zero mean and variance $\sigma^2$ \cite{matsuda2017multi}.

%%%%%%%%%%%%%%%%%%%%%%%%%%
\subsection{Transform-based Tensor Model}

Let $\mathcal{A} \in \mathbb{R}^{N_1 \times N_2 \times N_3}$ denote a third-order tensor. $\mathcal{A}(:,j,k)$,\; $\mathcal{A}(i,:,k)$,\; $\mathcal{A}(i,j,:)$ denote mode-1, mode-2, mode-3 tubes of $\mathcal{A}$, and $\mathcal{A}(:,:,k)$, $ \mathcal{A}(:,j,:)$, $\mathcal{A}(i,:,:)$ denote the frontal, lateral, and horizontal slices. The Frobenius norm of $\mathcal{A}$ is defined as $\norm{\mathcal{A}}_F=\sqrt{\sum_{i=1}^{N_1}\sum_{j=1}^{N_2}\sum_{k=1}^{N_3}\mathcal{A}_{ijk}^2}$. The operator vec($\cdot$) transforms tensors and matrices into vectors. Let $X^{T}$ and $\mathcal{A}^{\dagger}$ denote the transposes of a matrix and a tensor, respectively.

\begin{definition} \cite{liu2017fourth}
	Given an invertible discrete transform $\mathcal{L}:\mathbb{R}^{1\times 1 \times N_3}\rightarrow \mathbb{R}^{1\times 1 \times N_3}$, the elementwise multiplication $\circ$, and $\bm{a},\bm{b} \in \mathbb{R}^{1\times 1 \times N_3}$, the tubal-scalar multiplication is defined as
	\begin{equation}
	\bm{a}\bullet\bm{b}=\mathcal{L}^{-1}(\mathcal{L}(\bm{a})\circ\mathcal{L}(\bm{b})),
	\end{equation} 
	where $\mathcal{L}^{-1}$ is the inverse of $\mathcal{L}$.
\end{definition}

\begin{definition} \cite{liu2017fourth}
	The $\mathcal{L}$-product $\mathcal{A}=\mathcal{B} \, \bullet \, \mathcal{C}$ of $\;\mathcal{B}\in\mathbb{R}^{N_1\times r \times N_3}$ and $\mathcal{C}\in\mathbb{R}^{r\times N_2 \times N_3}$ is a tensor of size $N_1 \times N_2 \times N_3$, $\mathcal{A}(i,j,:)=\sum_{s=0}^{r}\mathcal{B}(i,s,:)\bullet\mathcal{C}(s,j,:)$, for $i\in[N_1] $ and $ j\in[N_2]$.
\end{definition}

\begin{definition} \cite{liu2017fourth}
	 The transpose of $\mathcal{A}$, $\mathcal{A}^\dagger \in \mathbb{R}^{N_2\times N_1 \times N_3}$ satisfies $\mathcal{L}\left( \mathcal{A}^\dagger\right) ^{(i)}=\left( \mathcal{L}(\mathcal{A})^{(i)}\right) ^T$, $i\in [N_3]$.
\end{definition}
	
\begin{definition} \cite{liu2017fourth}
	Identity tensor based on $\mathcal{L}$-product is defined as $\mathcal{I}\in\mathbb{R}^{N_1\times N_1 \times N_2} $ with $\mathcal{L}\left( \mathcal{I}\right) ^{(i)}$, $i\in [N_2]$ are $N_1 \times N_1 $ identity matrices.
\end{definition}

\begin{definition} \cite{liu2017fourth}
	$\mathcal{A}$ is $\mathcal{L}$-orthogonal if $\mathcal{A}\bullet\mathcal{A}^{\dagger}=\mathcal{A}^{\dagger}\bullet\mathcal{A}=\mathcal{I}$.
\end{definition}

\begin{definition} \cite{liu2017fourth}
	The transform domain singular value decomposition $\mathcal{L}$-SVD of $\mathcal{A} \in\mathbb{R}^{N_1 \times N_2 \times N_3} $ is given by $\mathcal{A}=\mathcal{U}\bullet\mathcal{S}\bullet\mathcal{V}^{\dagger}$, where $\mathcal{U}$ and $\mathcal{V}$ are $\mathcal{L}$-orthogonal tensors of size $N_1\times N_1\times N_3$ and $N_2\times N_2\times N_3$ respectively, and $\mathcal{S}$ is a diagonal tensor of size $ N_1 \times N_2\times N_3$. The entries of $\mathcal{S}$ are called the singular values of $\mathcal{A}$, and the number of non-zero ones is called the $\mathcal{L}$-rank of $\mathcal{A}$.
\end{definition}

% \begin{definition} \cite{liu2017fourth}
% 	For $\mathcal{A}\in \mathbb{R}^{N_1 \times N_2 \times N_3}$ with $\mathcal{L}$-rank of $r$, the $r$-dimensional tensor-column subspace $\mathcal{S}$ spanned by the columns of $\mathcal{A}$ is defined as
% 	\begin{equation}
% 	\mathcal{S}=\left\lbrace \mathcal{X}|\mathcal{X}=\mathcal{A}_1\bullet\bm{c}_1 +\mathcal{A}_2\bullet \bm{c}_2+\cdots+\mathcal{A}_{N_2}\bullet \bm{c}_{N_2}\right\rbrace,
% 	\end{equation}
% 	where $\bm{c}_j,j\in{[N_2]}$, is an arbitrary tubal scalar of length $N_3$.
% \end{definition}

% \begin{remark}
% 	If $\mathcal{S}$ is spanned by the columns of $\mathcal{A}\in\mathbb{R}^{N_1 \times N_2 \times N_3}$, $\mathcal{P}\triangleq\mathcal{A}\bullet(\mathcal{A}^{\dagger}\bullet\mathcal{A})^{-1}\bullet\mathcal{A}^{\dagger}$ is an orthogonal projection onto $\mathcal{S}$ when $\mathcal{A}^{\dagger}\bullet\mathcal{A}$ is invertible.
% \end{remark}

 In order to verify the validity of the model, we use an IKEA 3D chair model to generate a ground truth tensor of size $60\times 60\times 15$ (details are given in Section \ref{subsec:Data}), then the tensor is transformed to its frequency domain by fast fourier transformation (FFT) and discrete cosine transformation (DCT). Fig. 2 shows the empirical cumulative distribution function (CDF) of the tensor singular values. For $\mathcal{L}$-SVD with FFT, 3 out of 60 singular values capture 95\% of the energy, while the corresponding number of singular values are 38 and 54 for matrix-SVD and traditional tensor CP-Decomposition, respectively. For $\mathcal{L}$-SVD with DCT, 4 singular values capture 95\% energy. Therefore, the low $\mathcal{L}$-rank property of transform-based tensor model is more appropriate for the RF tomography imaging problem than other methods; a similar conclusion is given in the context of fingerprint localization \cite{Liu2016Adaptive}.
%%%%%%%%%%%%%%%%%%%%%%%%%%
\begin{figure}[t]
	\centering
	\includegraphics[totalheight=5.5cm]{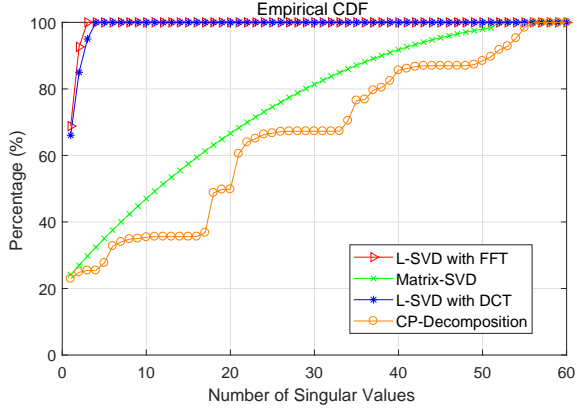}
	\caption{The CDFs of singular values of $\mathcal{L}$-SVD with FFT, $\mathcal{L}$-SVD with DCT, matrix-SVD and CP-Decomposition.}
	\label{fig2}
	\vspace{-0.17in}
\end{figure}

\subsection{Problem Formulation}

Let $K$ be the number of nodes within the network, then the total number of two-way links we can obtain is $S=(K^2-K)/2$. Assuming that we implement $M \ (1\le M\le S)$ times measurements, and each measurement involves a unique link. Those $M$ nodes pairs are indexed as $(v_{i_m},v_{j_m}),\ m\in[M]$. We obtain the following linear measurement:
\begin{equation}\label{equa:model}
\begin{split}
& \bm{y}_m  = \bm{P}_{t}-\bm{P}_{i_mj_m}-\overline{\bm{P}}(d_{i_m,j_m})\\
&=\sum_{n_1,n_2,n_3}{\mathcal{D}_{i_mj_m}(n_1,n_2,n_3)\mathcal{X}(n_1,n_2,n_3)} + \bm{Z}_{i_mj_m}^{\left( 2\right)},
\end{split}
\end{equation}
where $\bm{y}_m$ is referred to as the $m$-th measured total fading loss. Then we stack these RF signal measurements $\bm{y}_m$ into a measurement vector $\bm{y}\in\mathbb{R}^{M}$ \cite{kong2014data}. Given linear measurements $\bm{y}_m=\left\langle \mathcal{X},\mathcal{A}_m\right\rangle=\left( \text{vec}\left(\mathcal{A}_m \right)\right)^T \cdot \text{vec}\left(\mathcal{X} \right) ,1\leq m\leq M$ of a loss field tensor $\mathcal{X}\in\mathbb{R}^{N_1 \times N_2\times N_3}$ with $\mathcal{L}$-rank $r$ and the sensing tensors $\mathcal{A}_m\in  \mathbb{R}^{N_1 \times N_2\times N_3}$. With a linear map $\mathcal{H}(\cdot):\mathbb{R}^{N_1 \times N_2\times N_3}\rightarrow \mathbb{R}^{M}$ \cite{jain2013low},  (\ref{equa:model}) is rewritten as follows:

\begin{equation} 
\bm{y} = \mathcal{H}(\mathcal{X}) + \bm{w},
\end{equation}
%\begin{equation} 
%   \bm{y} = \begin{bmatrix}
%\bm{y}_1  \\
%\bm{y}_2  \\
%\vdots\\
%\bm{y}_d  \\
%\end{bmatrix}
%=\begin{bmatrix}
%\left\langle \mathcal{X},\mathcal{A}_1\right\rangle  \\
%\left\langle \mathcal{X},\mathcal{A}_2\right\rangle  \\
%\vdots\\
%\left\langle \mathcal{X},\mathcal{A}_d\right\rangle  \\
%\end{bmatrix}+
%\begin{bmatrix}
%\bm{w}_1  \\
%\bm{w}_2  \\
%\vdots\\
%\bm{w}_d  \\
%\end{bmatrix},
%\end{equation}
where $\bm{w}=(\bm{Z}_{i_1j_1}^{\left( 2\right)},\bm{Z}_{i_2j_2}^{\left( 2\right)},\cdots,\bm{Z}_{i_Mj_M}^{\left( 2\right)})^T$ denotes the noise vector.

The goal of RF tomographic imaging is to recover the loss field tensor $\mathcal{X}$ from the measurement vector $\bm{y}$.
We formulate this problem as a low $\mathcal{L}$-rank tensor sensing problem: 
\begin{equation}\label{equa:TS}
\begin{aligned}
{\widehat{\mathcal{X}}} = &\argmin_{\mathcal{X}\in  \mathbb{R}^{N_1 \times N_2\times N_3}} ~~\norm{\bm{y}-\mathcal{H}(\mathcal{X})}_{F}^2, \\
&\text{s.t.}\ \ \mathrm{rank}(\mathcal{X})\leq r.
\end{aligned}
\end{equation}

\begin{algorithm}[t] 
	\caption{Alt-Min: $\mathrm{AM}(\mathcal{H}(\cdot),\bm{y},r,L)$} 
	\label{alg:Tubal-Alt-Min-Sense} 
	\begin{algorithmic}[1] 
		\REQUIRE  linear map $\mathcal{H}(\cdot)$, measurement vector $\bm{y}$, $\mathcal{L}$-rank $r$, iteration number $L$. 
		\STATE Initialize ${\mathcal{U}}^{0}$ randomly;
		\FOR{$\ell=1$ to $L$} 
		\STATE ~~~$\mathcal{V}^{\ell}~\leftarrow~\mathrm{LS}(\mathcal{H}(\cdot),\mathcal{U}^{\ell-1},\bm{y},r)$;
		\STATE ~~~$\mathcal{U}^{\ell}~\leftarrow~\mathrm{LS}(\mathcal{H}(\cdot),\mathcal{V}^{\ell}\quad, \bm{y},r)$;
		\ENDFOR
		\ENSURE Pair of tensors $(\mathcal{U}^{L},\mathcal{V}^{L})$.
	\end{algorithmic}
	\vspace{-0.07in}
\end{algorithm}

\begin{algorithm}[t] 
	\caption{Least Squares Minimization: $\mathrm{LS}(\mathcal{H}(\cdot),\mathcal{U},\bm{y},r)$} 
	\label{alg:Least Squares Update} 
	\begin{algorithmic}[1] 
		\REQUIRE  linear map $\mathcal{H}(\cdot)$, tensor $\mathcal{U}\in \mathbb{R}^{N_1\times r \times N_3}$, measurement vector $\bm{y}$, $\mathcal{L}$-rank $r$.
		\STATE $\mathcal{V} = \argmin\limits_{\mathcal{V}\in \mathbb{R}^{r \times N_2 \times N_3}} ~~\norm{\bm{y}-\mathcal{H}(\mathcal{U}\bullet\mathcal{V})}_{F}^2$;
		\ENSURE tensor $\mathcal{V}$
	\end{algorithmic}
	\vspace{-0.07in}
\end{algorithm}

%%%%%%%%%%%%%%%%%%%%%%%%%%%%%%%%%%%%%%%

\section{Solution Algorithm Via Alt-Min}\label{sec:solution}
In this section, we present a novel iterative algorithm, called Alt-Min, and review its optimization and implementation.

%%%%%%%%%%%%%%%%%%%%%%%%%%%%%%
\subsection{The Alt-Min Algorithm}\label{subsec:imp}

To enable alternating minimization, we represent the loss field tensor as the $\mathcal{L}$-product of two smaller tensors \cite{liu2016low}, i.e., ${\mathcal{X}}  = \mathcal{U}\bullet\mathcal{V}$, $\mathcal{X}\in \mathbb{R}^{N_1 \times N_2 \times N_3}$, $\mathcal{U}\in \mathbb{R}^{N_1 \times r \times N_3}$ and $\mathcal{V}\in \mathbb{R}^{r \times N_2 \times N_3}$. Then we reformulate equation (\ref{equa:TS}) as the following non-convex optimization problem:
 \begin{equation}\label{obs_model1}
{\widehat{\mathcal{X}}} =\argmin_{\mathcal{U}\in\mathbb{R}^{N_1 \times r \times N_3},\mathcal{V}\in\mathbb{R}^{r \times N_2 \times N_3}} ~~\norm{\bm{y}-\mathcal{H}(\mathcal{U}\bullet\mathcal{V})}_{F}^2.
\end{equation}

The main idea of Alt-Min is to iteratively estimate two low $\mathcal{L}$-rank tensors $\mathcal{U}$ and $\mathcal{V}$, each of $\mathcal{L}$-rank $r$. The key step  is least squares (LS) minimization (see Alg. \ref{alg:Least Squares Update}). The detailed implementation of LS minimization are given below.

We adopt circulant algebra \cite{liu2016low, gleich2013power} to extend matrix algebra to third-order tensors. A tubal scalar represents a vector of length $N_3$, and the corresponding space is denoted as $\mathbb{K}$. Let $\mathbb{K}^{N_1 \times N_2}$ denote the space of $N_1 \times N_2$ tubal matrices where each element is a tubal scalar in $\mathbb{K}$. Let $\underline{\alpha}$ be a tubal scalar, and $\underline{\bm{A}}$ be a tubal matrix. We use the operator $\text{circ}(\cdot)$ \cite{gleich2013power} to map circulants to their corresponding circular matrices, which are tagged with the superscript $c$, i.e., $\underline{\alpha}^{c},\underline{\bm{A}}^{c}$:
\begin{equation*}
\underline{\alpha}^c = \text{circ}(\underline{\alpha})=
\begin{bmatrix}
\alpha_1 &  \alpha_{N_3}   & \cdots & \alpha_2    \\
\alpha_2 &  \alpha_1   &  \cdots    &  \cdots  \\
\vdots & \vdots &\vdots & \alpha_{N_3}      \\
\alpha_{N_3}      & \alpha_{N_3-1} & \cdots & \alpha_1
\end{bmatrix},
\end{equation*}
\begin{equation*}
\underline{\bm{A}}^c = \text{circ}(\underline{\bm{A}})=
\begin{bmatrix}
\text{circ}(\underline{\bm{A}}_{1,1}) &  \cdots & \text{circ}(\underline{\bm{A}}_{1,N_2})   \\
\vdots &    \vdots    &  \vdots  \\
\text{circ}(\underline{\bm{A}}_{N_1,1})      &  \cdots & \text{circ}(\underline{\bm{A}}_{N_1,N_2})
\end{bmatrix}.
\end{equation*}

For simplicity, we use $\bm{A}^c$ to represent the circular matrix of tensor $\mathcal{A}$. Then the $\mathcal{L}$-product $\mathcal{X}= \mathcal{U}\bullet\mathcal{V}$ has an equivalent matrix-product as:
\begin{equation}
%\mathcal{X}= \mathcal{U}*  \mathcal{V} \quad\leftrightarrow\quad 
\bm{X}^c=\bm{U}^c\bm{V}^c,
\end{equation}
where $\bm{X}^c \in \mathbb{R}^{N_1N_3 \times N_2N_3},  \ \bm{U}^c \in \mathbb{R}^{N_1N_3\times rN_3}, \ \bm{V}^c \in \mathbb{R}^{rN_3 \times N_2N_3}$. We can transform the LS minimization in Alg. 2 to the corresponding circular matrix representation:
\begin{equation}
\widehat{\bm{V}}^c=\argmin_{ \bm{V}^c \in \mathbb{R}^{rN_3 \times N_2N_3}} ~~\norm{\bm{y}-\mathcal{H}^c(\bm{U}^c\bm{V}^c)}_{F}^2,
\end{equation}
where $\mathcal{H}^c(\cdot):  \mathbb{R}^{N_1N_3 \times N_2N_3} \to  \mathbb{R}^{M}$  is the corresponding linear map in the circular matrix representation, with $\bm{y}=\mathcal{H}^c(\bm{X}^c)$. Each sensing tensor $\mathcal{A}_m$ is transformed into its circular matrix $\bm{A}_m^c \in \mathbb{R}^{N_1N_3 \times N_2N_3}$, and $\bm{y}_m=\left\langle \bm{X}^c, {\bm{A}^c_m}\right\rangle,1\le m \le M $. Similarly, we can estimate $\bm{U}^c$ in the following way:
\begin{equation}
\widehat{\bm{U}}^c=\argmin_{ \bm{U}^c \in \mathbb{R}^{N_1N_3 \times rN_3}} ~~\norm{\bm{y}-{\mathcal{H}^c}^T({\bm{V}^c}^T{\bm{U}^c}^T)}_{F}^2.
\end{equation}

We perform the following steps to solve this non-convex optimization problem:

\textbf{Step 1)}. $\bm{U}^c$ is used to form a block diagonal matrix $\bm{B}_1$ of size $N_1N_2N_3^2\times rN_2N_3^2$, and the number of $\bm{U}^c$ is $N_2N_3$,
\begin{equation}
\bm{B}_1=
\begin{bmatrix}
\bm{U}^c &   &  &    \\
&  \bm{U}^c  &    &    \\
& &\ddots &       \\
& &  &\bm{U}^c 
\end{bmatrix}.
\end{equation}

\textbf{Step 2)}. Stack all the columns of $\bm{V}^c$, and then $\bm{V}^c$ is vectorized to a vector $\bm{b}$ of size $rN_2N_3^2\times 1$ as follows:
\begin{equation}
\begin{split}
&\bm{b}=\text{vec}(\bm{V}^c) \\
&=[\bm{V}^c(:,1)^T,\; \bm{V}^c(:,2)^T,\; \ldots,\; \bm{V}^c(:,N_2N_3)^T]^T.
\end{split}
\end{equation}

\textbf{Step 3)}. Each $\bm{A}_m^c,\;1\le m\le M$ is represented as a vector $\bm{c}_m$ of size $N_1N_2N_3^2\times1$ in the following way:
\begin{equation}
\begin{split}
&\bm{c}_m=\text{vec}(\bm{A}_m^c)\\
&=[\bm{A}_m^c(:,1)^T,\; \bm{A}_m^c(:,2)^T,\; \ldots,\;\bm{A}_m^c(:,N_2N_3)^T]^T,
\end{split}
\end{equation}
and then all the $\bm{c}_m$ are transformed into a matrix $\bm{B}_2$ of size $M \times N_1N_2N_3^2$:
\begin{equation}
\bm{B}_2=[\bm{c}_1,\; \bm{c}_2,\; \ldots,\; \bm{c}_M]^T.
\end{equation}

Therefore, the estimation of $\bm{V}^c$ is transformed into the following standard least squares minimization problem:
\begin{equation}
\widehat{\bm{b}}=\argmin_{\bm{b} \in \mathbb{R}^{rN_2N_3^2\times 1}} ~~\norm{\bm{y}-\bm{B}_2\bm{B}_1\bm{b}}_{F}^2.
\end{equation}

  \begin{figure*}[t]
 	\centering
 	\subfloat[ ]{%
 		\includegraphics[width=.22\textwidth]{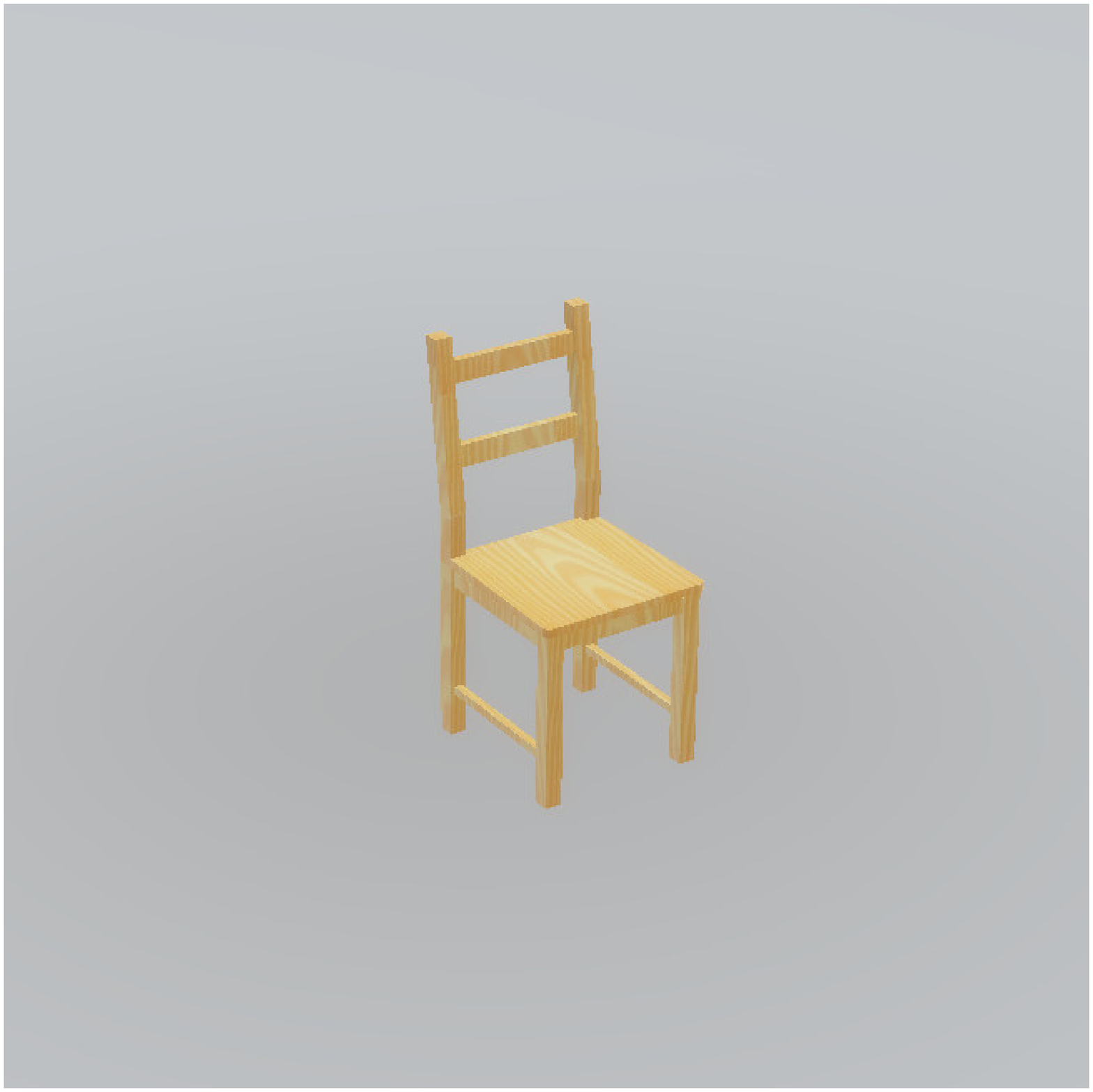}}\hspace{1em}%
 	\subfloat[ ]{%
 		\includegraphics[width=.22\textwidth]{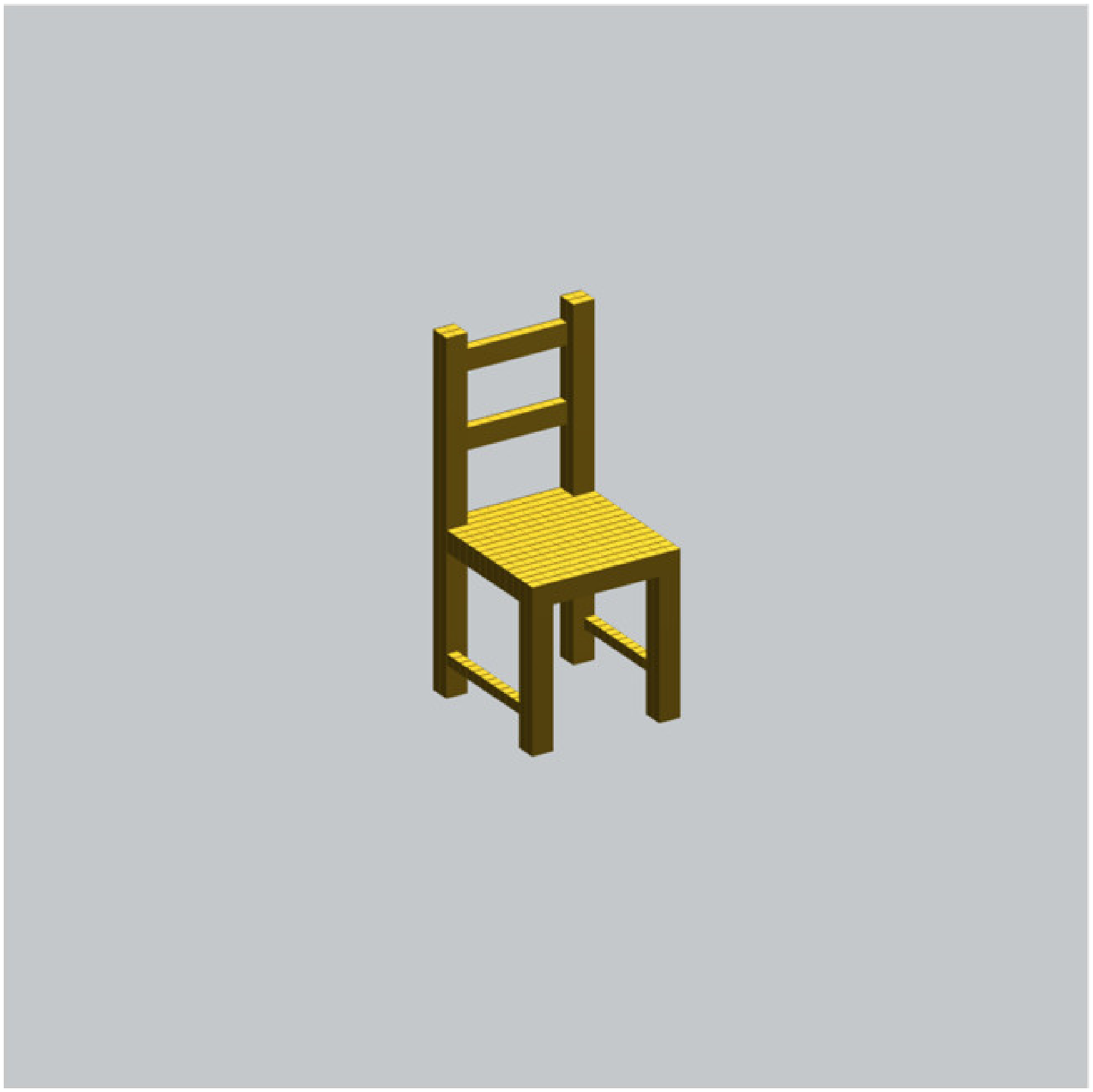}}\hspace{1em}%
 	\subfloat[ ]{%
 		\includegraphics[width=.22\textwidth]{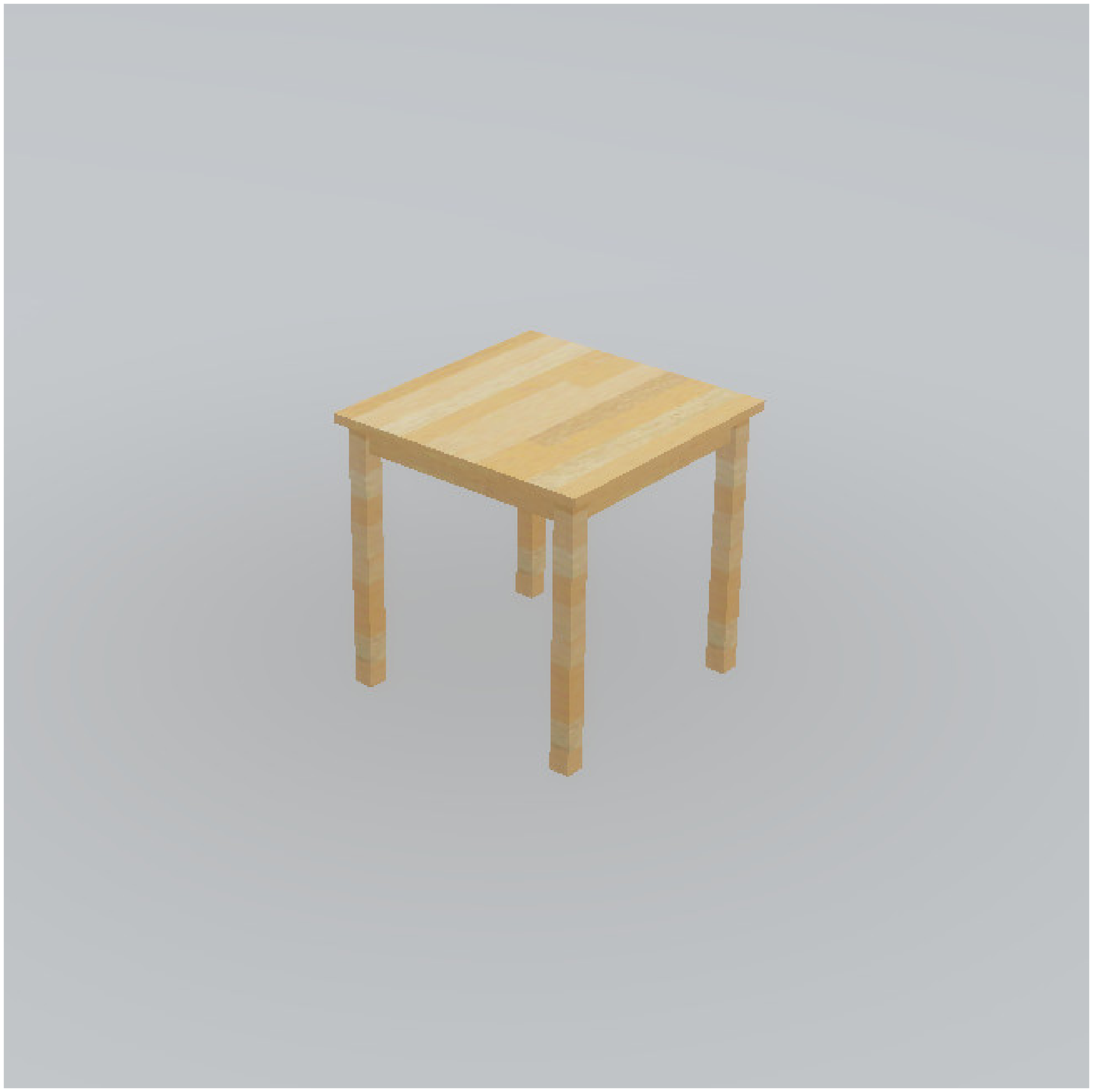}}\hspace{1em}%
 	\subfloat[ ]{%
 		\includegraphics[width=.22\textwidth]{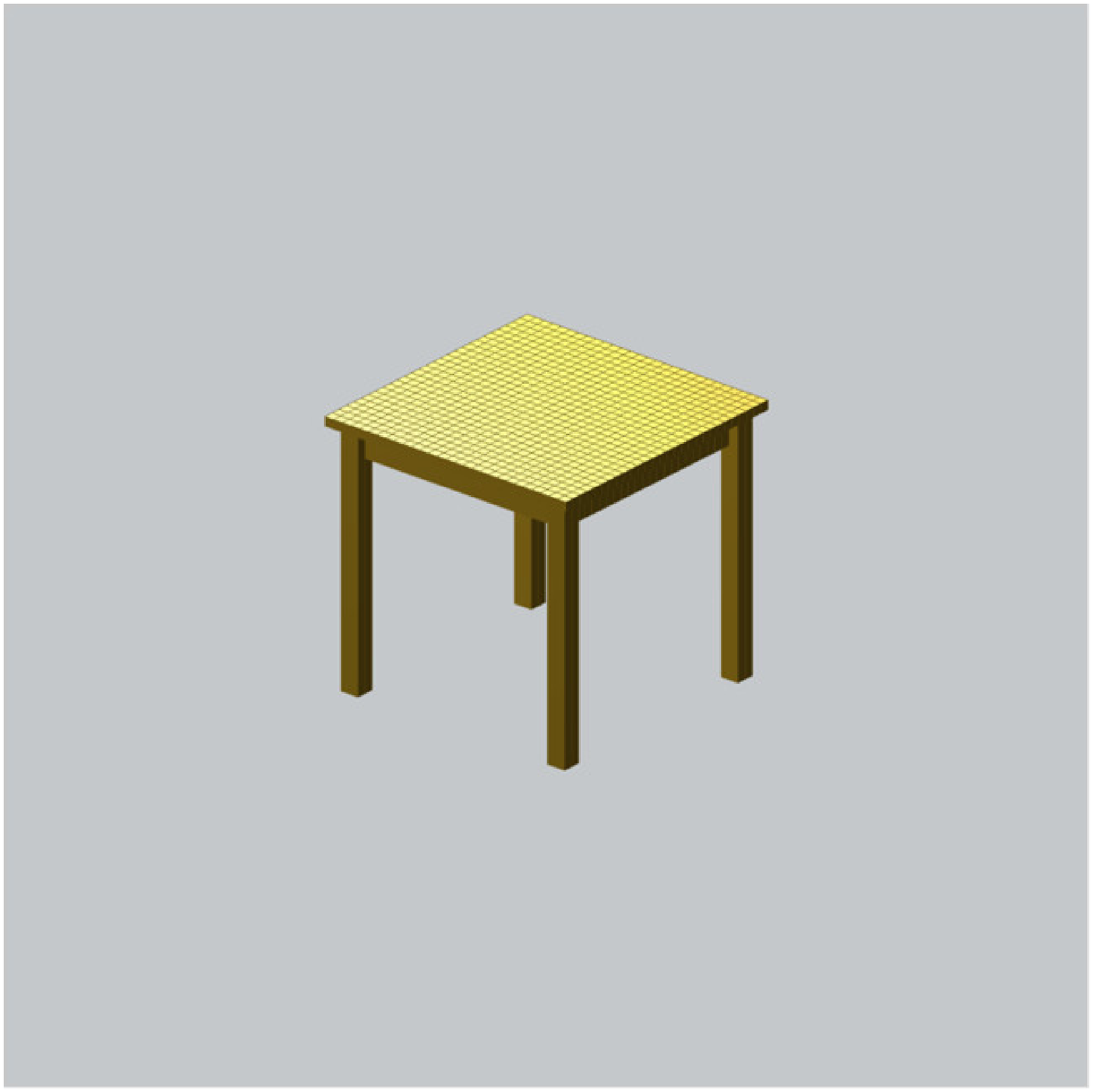}}\\
 	\caption{(a) and (c) are the 3D visualizations of two IKEA models, (b) and (d) are the corresponding recovery results.}
 	\vspace{-0.2in}
 \end{figure*}
 
%%%%%%%%%%%%%%%%%%%%%%%%%%%%%%%%%%%%%%%
\subsection{Algorithm Optimization}
\label{sect:alg_optimization}

The proposed Alt-Min requires large memory consumption and high computational-complexity. We propose improvements to resolve such problems.

\subsubsection{Optimization of Alt-Min}
As stated above, the loss field tensor and the sensing tensor are transformed to an unknown circular matrix $\bm{X}^c\in \mathbb{R}^{N_1N_3 \times N_2N_3}$ and a sensing circular matrix ${\bm{A}}_m^c \in \mathbb{R}^{N_1N_3 \times N_2N_3}$, respectively. The unknown circular matrix $\bm{X}^c$ consists of $N_1N_2N_3^2$ entries and if we set the sampling rate as 50\%, the total number of all sensing circular matrices is 0.5$N_1N_2N_3^2$. In this case, the space complexity of all sensing matrices is $O(N_1^2N_2^2N_3^4)$, and it is obvious that the memory requirement increases exponentially with the size of the tensor. To alleviate this problem, we propose a modified version of the implementation.
 
In circulant algebra, it is obvious that the first column of $\underline{\alpha}^c$ already contains all the entries of itself, and there is no need to recover the redundant information. For recovering the loss field tensor $\mathcal{X}$, we only need to recover the first column of each $\text{circ}(\underline{\bm{X}}_{i,j})$, which we set as the $i$-th tube of the $j$-th lateral slice: $\mathcal{X}(i,j,:)$. We use the Matlab function $\text{squeeze}(\cdot)$ to get a new definition:
\begin{equation*}
{\bm{X}}^s =
\begin{bmatrix}
\text{squeeze}(\mathcal{X}(1,1,:)) &  \cdots &  \text{squeeze}(\mathcal{X}(1,N_2,:))   \\
\vdots &    \ddots   &  \vdots  \\
\text{squeeze}(\mathcal{X}(N_1,1,:))      &  \cdots & \text{squeeze}(\mathcal{X}(N_1,N_2,:))
\end{bmatrix},
\end{equation*}
where $\text{squeeze}(\mathcal{X}(i,j,:))$ transforms the $i$-th tube of the $j$-th lateral slice of $\mathcal{X}$ 
into a vector of size $N_3\times1$.

We use the notation $\Leftrightarrow$ to denote a new mapping for $\mathcal{L}$-product as follows:
\begin{equation}
\mathcal{X}= \mathcal{U} \bullet \mathcal{V} \quad\Leftrightarrow\quad \bm{X}^s=\bm{U}^c\bm{V}^s,
\end{equation}
where $\bm{X}^s \in \mathbb{R}^{N_1N_3 \times N_2},  \ \bm{U}^c \in \mathbb{R}^{N_1N_3\times rN_3}, \ \bm{V}^s \in \mathbb{R}^{rN_3 \times N_2}$. We can transform the LS minimization in Alg. 2 to the following representation:
\begin{equation}
\widehat{\bm{V}}^s=~~\argmin_{ \bm{V}^s \in \mathbb{R}^{rN_3 \times N_2}} ~~\norm{\bm{y}-\mathcal{H}^s(\bm{U}^c\bm{V}^s)}_{F}^2,
\end{equation}
where $\mathcal{H}^s(\cdot):  \mathbb{R}^{N_1N_3 \times N_2} \to  \mathbb{R}^{M}$  is the corresponding linear map, with  $\bm{y}=\mathcal{H}^s(\bm{X}^s),\  \bm{y}_m=\left\langle \bm{X}^s, \bm{A}_m^s\right\rangle,1\le m \le M $. Similarly, we can estimate $\bm{U}^c$ in the following way:
\begin{equation}
\widehat{\bm{U}}^c=~~\argmin_{ \bm{U}^c \in \mathbb{R}^{N_1N_3 \times rN_3}} ~~\norm{{\bm{y}-\mathcal{H}^s}^T({\bm{V}^s}^T{\bm{U}^c}^T)}_{F}^2.
\end{equation}

\subsubsection{Complexity Analysis}

As stated above, if we use 50\% sampling rate, the original space complexity is $O(N_1^2N_2^2N_3^4)$. In the modified version, we transform $\mathcal{X},\ \mathcal{A}_m$ to $\bm{X}^s \in \mathbb{R}^{N_1N_3 \times N_2}, \ \bm{A}_m^s \in \mathbb{R}^{N_1N_3 \times N_2}$, and the space complexity decreases to $O(N_1^2N_2^2N_3^2)$, which is $1/N_3^2$ of the former value.
Note that we only calculate the reduction of space complexity for the sensing matrices, but there are additionally large reduction in intermediate variables. Therefore, the above algorithm optimization is a key enabler in our approach for inferring large three-dimensional physical space.  

%%%%%%%%%%%%%%%%%%%%%%%%%%%%%%%%%

\section{Performance Evaluation}

%%%%%%%

\subsection{Data Sets And Model Verification} \label{subsec:Data}
We compare the proposed algorithm Alt-Min with tensor-based compressed sensing \cite{matsuda2017multi} on the IKEA 3D datasets. We adopt an IKEA 3D chair model and table model to generate two ground truth tensors of the same size $60\times 60\times 15$. The $\mathcal{L}$-rank of the chair model is 3, and that of the table model is 4. Each 3D model is placed in the middle of the ``tensor" and occupies a part of the space. In this task, we mainly focus on the location and outline information, while the texture and color information are ignored.

% Further more, we also implement our own experiment and collect another ground-truth measurement dataset to verify our approach. 

% In our own experiment, we choose a region of size $6\text{m} \times 6\text{m} \times 1.5\text{m}$, which is devided into $60\times60\times15$ voxels with each voxel of the same size $10\text{cm}\times 10\text{cm}\times 10\text{cm}$. Then a wireless peer-to-peer network consists of 144 wireless nodes is established, and the nodes are uniformly deployed on the sides of the region. We randomly choose a pair of wireless nodes to get one measurement value. After $M$ times simulation, we get an RSS vector $\bm{y} \in\mathbb{R}^{M}$ for the follow-up processing.
%%%%%%%
\begin{figure*}[t]
	\begin{multicols}{3}
		\includegraphics[width=6cm]{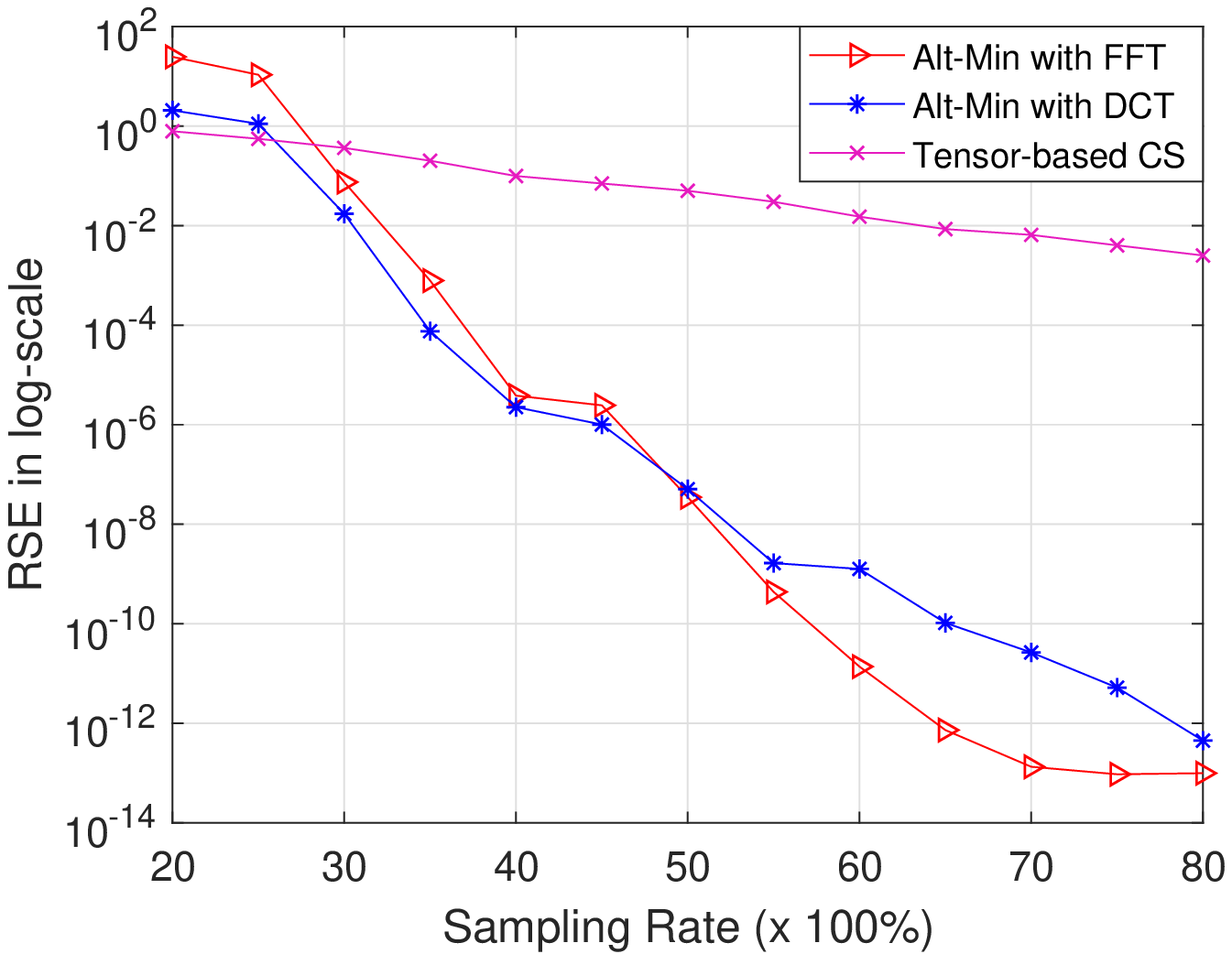}\par\caption{RSEs vs sampling rates.}
		\includegraphics[width=6cm]{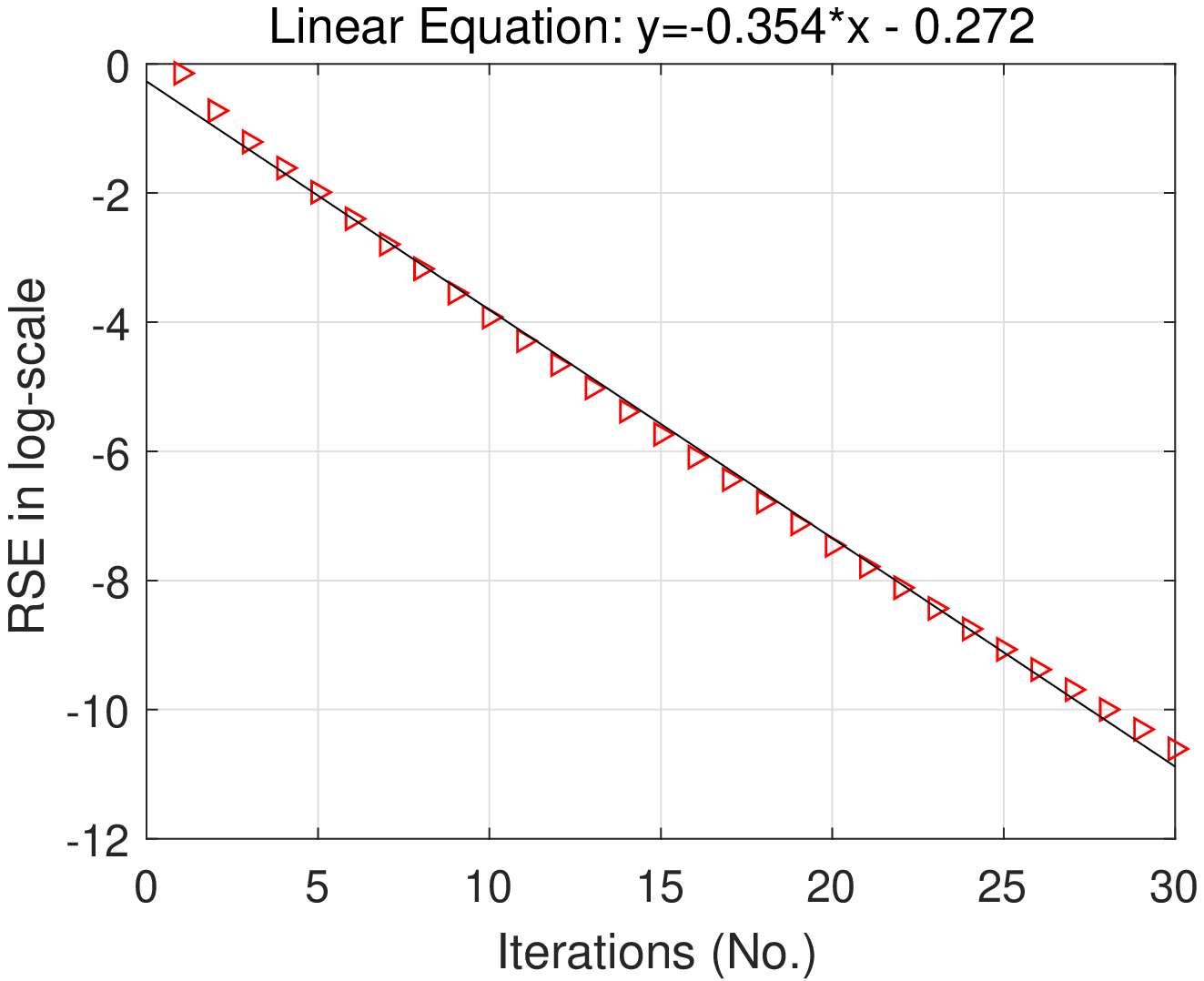}\par\caption{Alt-Min with FFT.}
		\includegraphics[width=6cm]{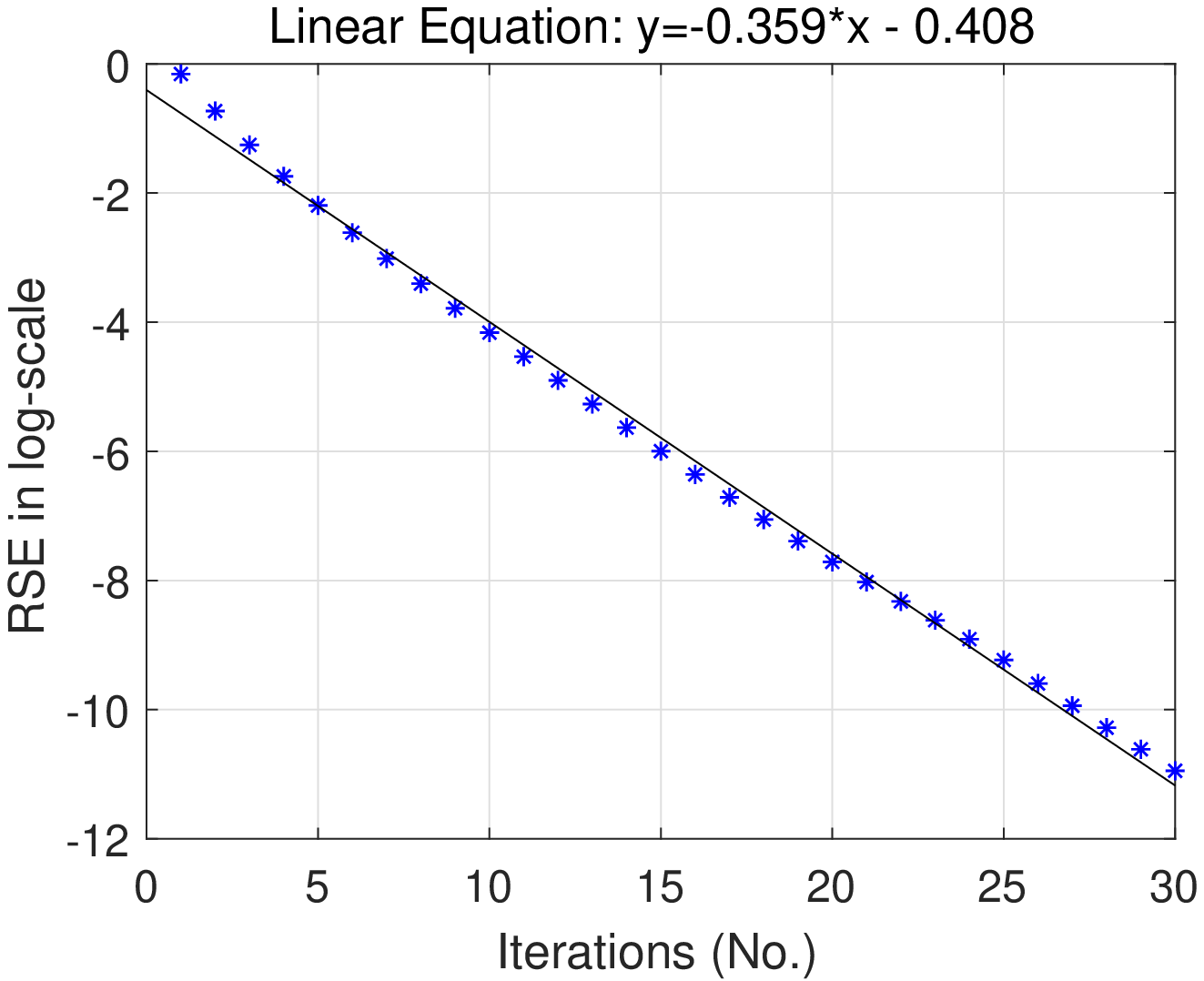}\par\caption{Alt-Min with DCT.}
	\end{multicols}
	\vspace{-0.3in}
\end{figure*}

%%%%%%%
\subsection{Algorithm Comparison Metrics}
We compare the Alt-Min against the recently proposed tensor-based compressed sensing \cite{matsuda2017multi} on the IKEA 3D datasets. Note that we carry out two versions of the Alt-Min: Alt-Min with FFT and Alt-Min with DCT. The tensor-based compressed sensing uses TNN as the regularization, in which the t-SVD is conducted in every iteration. Our algorithm is based on the bi-linear factorization, and we only need to iteratively update the two smaller tensors. For quantitative comparison, we adopt two metrics: the recovery error and the convergence speed.
\begin{itemize}
	\item For recovery error, we use the metric relative square error, defined as $\text{RSE}=||\widehat{\mathcal{X}}-\mathcal{X}||_F/||{\mathcal{X}}||_F$.
	\item For the convergence speed, we linearly fitting the measured RSEs across iterations, and then compare the decreasing rate of each method.
\end{itemize}
%%%%%%%

\subsection{Performance Results}
Fig. 3 shows the 3D visualizations of an IKEA chair, an IKEA table, and our corresponding recovery results using Alt-Min with DCT. For all methods, we use 50\% sampling rate, and the maximum iteration number is set to 20. The final recovery error (RSE in log-scale) of the chair model is $10^{-8}$ in magnitude, and that of the table model is $10^{-7}$. We can observe that Alt-Min with DCT successfully recovers the outlines of two models. Note that we focus on the outlines of the models instead of the whole space, and the recovery results are artificially coloured for a better visualization. The following analysis is based on the experiments of the chair model.

To examine recovery error performance, we fix the maximum iteration numbers of three methods to be 20. Then we vary the sampling rate from 20\% to 80\% by selecting wireless links randomly \cite{liu2015cdc}. Each sampling rate is measured 5 times, and the average recovery error results are computed. Fig. 4 depicts the RSEs of Alt-Min with FFT, Alt-Min with DCT and tensor-based compressed sensing for varying sampling rates. For low sampling rates (20\% $\sim$ 25\%), tensor-based compressed sensing performs better than Alt-Min with FFT and Alt-Min with DCT. For sampling rates varying from 30\% to 80\%, the RSEs of two Alt-Min methods decrease significantly, while that of tensor-based compressed sensing decreases very slowly. For 80\% sampling rate, the RSE of Alt-Min with FFT is around $10^{-13}$ in magnitude, and that of Alt-Min with DCT is a little higher, while the the RSE of tensor-based compressed sensing is at around $10^{-3}$. 

Fig. 5 and Fig. 6 show the convergence rates of Alt-Min with FFT and with DCT, respectively. The sampling rate is fixed to be $50\%$, the maximum iteration number is set at 30. A linear fit to the data is also shown. Note that if we set the RSE threshold to be $10^{-10}$, we only need approximately 27 iterations for both methods.

%%%%%%%%%%%%%%%%%%%%%%%%%%%%%%%%%%%%%%%
\section{Conclusion}

In this paper, we use the transform-based tensor model to formulate the RF tomographic imaging as a tensor sensing problem, which can fully exploit the geometric structures of the three-dimensional loss field tensor. Then we propose a fast iterative algorithm Alt-Min for the low $\mathcal{L}$-rank tensor sensing. The loss field tensor is factorized as the $\mathcal{L}$-product of two smaller tensors, and then Alt-Min alternately estimates those two tensors by LS minimization. The evaluation results on IKEA 3D datasets have demonstrated that Alt-Min significantly improves the recovery error and convergence speed compared to prior tensor-based compressed sensing.

\bibliographystyle{IEEEbib}
\bibliography{icme2018template}

\end{document}